\documentclass{article}
\usepackage{arxiv}
\usepackage[utf8]{inputenc} 
\usepackage[T1]{fontenc}    
\usepackage{hyperref}       
\usepackage{url}            
\usepackage{booktabs}       
\usepackage{amsfonts}       
\usepackage{nicefrac}       
\usepackage{microtype}      
\usepackage{lipsum}
\usepackage{graphicx}
\graphicspath{ {./images/} }
 \usepackage{dcolumn}
 \usepackage{endnotes}
 \usepackage{amssymb}
\usepackage{caption}
 \usepackage{setspace}
 \usepackage{array}
  \usepackage{float} 
\usepackage{geometry}
 \usepackage{multirow}
 \usepackage{pdfpages}
\usepackage{amsmath,amssymb}
\usepackage{arydshln}
 \usepackage{mathabx}
 \usepackage{comment}
 \usepackage{longtable}
\usepackage{tabularray}
\usepackage{subcaption}

\title{Two-Phase Segmentation Approach for Accurate Left Ventricle Segmentation in Cardiac MRI using Machine Learning}

\author{
 Maria Tamoor \\
  Forman Christian College, \\
  Lahore, Pakistan \\
  \texttt{mariatamoor@fccollege.edu.pk} \\
   \And
   Abbas Raza Ali \\
Citi Innovation Lab\\
London, United Kingdom\\
  \texttt{abbas.raza.ali@gmail.com} \\
 \And
 Philemon Philip \\
 Forman Christian College, \\
  Lahore, Pakistan \\
  \texttt{philemonphilip98@gmail.com} \\
  \And
  Ruqqayia Adil \\
  NSHS, NUST,\\
  Islamabad, Pakistan\\
  \texttt{ruquiyya@gmail.com} \\
  \And
  Rabia Shahid \\
  Government college university, \\
  Faisalabad, Pakistan\\
  \texttt{rabiashahid993@gmail.com} \\
  \And
  Asma Naseer \\
  NUCES\\
  Lahore, Pakistan\\
  \texttt{asma.naseer@nu.edu.pk} \\
}
\begin{document}
\maketitle
\begin{abstract}
Accurate segmentation of the Left Ventricle (LV) holds substantial importance due to its implications in disease detection, regional analysis, and the development of complex models for cardiac surgical planning. CMR is a golden standard for diagnosis of serveral cardiac diseases. LV in CMR comprises of three distinct sections: Basal, Mid-Ventricle, and Apical. This research focuses on the precise segmentation of the LV from Cardiac MRI (CMR) scans, joining with the capabilities of Machine Learning (ML). The central challenge in this research revolves around the absence of a set of parameters applicable to all three types of LV slices. Parameters optimized for basal slices often fall short when applied to mid-ventricular and apical slices, and vice versa. To handle this issue, a new method is proposed to enhance LV segmentation. The proposed method involves using distinct sets of parameters for each type of slice, resulting in a two-phase segmentation approach. The initial phase categorizes images into three groups based on the type of LV slice, while the second phase aims to segment CMR images using parameters derived from the preceding phase. A publicly available dataset (Automated Cardiac Diagnosis Challenge (ACDC)) is used. 10-Fold Cross Validation is used and it achieved a mean score of 0.9228. Comprehensive testing indicates that the best parameter set for a particular type of slice does not perform adequately for the other slice types. All results show that the proposed approach fills a critical void in parameter standardization through a two-phase segmentation model for the LV, aiming to not only improve the accuracy of cardiac image analysis but also contribute advancements to the field of LV segmentation.
\end{abstract}

In human body, each organ holds individual significance, but the heart is widely recognized as the most crucial component. The human heart comprises four primary chambers: the Left Ventricle (LV), Right Ventricle (RV), Left Atrium (LA), and Right Atrium (RA). Among these, the Left Ventricle (LV) plays a pivotal role, particularly contributing to the diagnosis of various cardiovascular diseases. LV does not only ciculates blood but holding the key to our heart’s well-being. The LV is also renowned for its remarkable power and often serves as the epicenter of cardiac diseases \cite{groenewegen2020epidemiology}. Heart failure occurs when the volume of blood pumped throughout the body is insufficient. This can happen when the heart either fails to adequately fill with the blood the body requires or lacks the strength to consistently circulate blood throughout the body. Therefore, the accurate detection of Left Ventricle (LV) function is of paramount importance to cardiologists. This helps them to identify any irregularities within the human heart and allows timely intervention. Various diagnostic methods, such as Magnetic Resonance Imaging (MRI), Computed Tomography (CT) scans, and echocardiography, are frequently employed to facilitate the early detection of heart diseases. Accurate left ventricular (LV) segmentation from Cardiac MRIs (CMR) supports healthcare professionals in the early detection of cardiac diseases, contributing to the potential saving of numerous lives \cite{tamoor2021two-stage}. CMR imaging results in a three-dimensional (3-D) representation of the heart. To facilitate segmentation, this 3-D image is typically converted into multiple two-dimensional (2-D) slices. Deviation of the size and structure of LV from normal parameters is a primary indicator for detecting several heart related diseases. For instance, the contraction and thickening of the left ventricle significantly contribute to evaluating insufficient blood supply to the cardiac tissue (Ischemia)\cite{bernard2018deep}. In contrast, a decrease in left ventricular output or ejection fraction may represent a late complication associated with elevated vascular resistance (Hypertension) \cite{bernard2018deep,yang2017left}.

Automated segmentation of the Left Ventricle (LV) faces several significant challenges \cite{hajiaghayi20183d}. These challenges include weak boundaries, intensity homogeneity, intensity leakage, artifacts, presence of outflow track, intensity overlap, low resolution and contrast \cite{yang2017left,wang2013active}. Structural variations across multiple slices (Basal/LVOT (Left Ventricle Overflow Tract), Mid-Ventricle, and Apical \cite{forte2019normal}) are also an issue while segmenting LV accurately. The Basal region, situated at the top of the LV, exhibits an oval shape rather than a perfect circle. In contrast, the Mid-Ventricle assumes a donut-like structure. And the Apical portion is the bottom most part of the LV, is notably small, sometimes even invisible to the eye. Furthermore, the heart undergoes two fundamental phases during its cardiac cycle. One is the diastolic phase and the other is the systolic phase. The diastolic phase marks the moment when blood is actively pumped in. In contrast, the systolic phase occurs when the heart expels all its blood \cite{galvez-pol2020active}. Despite these challenges, reliable and robust segmentation methods should be adopted to overcome these issues \cite{moradi2019mfp-unet}.

CMR employs a powerful magnetic field and radio waves to generate highly detailed images of the heart \cite{pennell2006myocardial}, offering a clear view of its internal structure. Importantly, this imaging technique does not expose patients to any harmful radiation \cite{symons2019cmr}. 
The task under study is to accurately segment out the LV from CMR scans via Machine Learning (ML). ML depicts the system’s capability to learn something from a collection data to structure process of creating a model to solve a task \cite{janiesch2021machine} \cite{vale2021efficient}. ML allows a system to develop a skill to solve a problem \cite{medar2017impact}. Various ML techniques are available for image segmentation, including Active Contour Model (ACM), Markov Random Field (MRF), Kernel Support Vector Machine (KSVM), and more \cite{seo2020machine}. One notably effective method for this purpose is Active Contour Models (ACM) \cite{pratondo2017integrating}. Typically, supervised learning techniques, which require knowledge of the correct output for training, are computationally expensive and time-consuming \cite{navab2016v-net}. In contrast, unsupervised learning techniques, which work without knowing the correct output, rely on guided image features to segment the desired parts of an image \cite{tamoor2021two-stage}. This makes unsupervised learning more robust, especially in the context of medical imaging \cite{aganj2018unsupervised}. Furthermore, Deep Learning (DL) techniques have proven highly effective in medical image segmentation \cite{hesamian2019deep}. However, DL methods have some drawbacks, including high memory requirements, significant computational costs, complexity, potential for overfitting, and long training times \cite{lan2018survey}. 

 As mentioned earlier, numerous methods have been introduced to achieve accurate left ventricular (LV) segmentation. However, these methods exhibit sensitivity to hyperparameters, necessitating extensive training and validation for proper tuning. Given the structural variations in the LV across different slices and phases within a single MRI, it becomes impractical to define a universal set of parameters that consistently produces optimal results for all types of slices. Optimized parameters for basal slices frequently yield suboptimal results for the other two types, and vice versa. This research gap underscores the critical need for an original approach to achieve accurate LV segmentation. To address this challenge, parameter Optimization to achieve precise LV segmentation is proposed. Different parameter sets are used when working with distinct types of LV slices. Hence, a two-phase methodology for LV segmentation is proposed in this paper. The first phase of the model involves the classification of CMR images. While the second phase is the actual segmentation of these images. Importantly, the choice of parameters is determined by the type of slice from the first phase. The central objective of this research is to demonstrate that employing different parameter sets for distinct types of slices can significantly impact the overall segmentation quality. 

 The main contributions of this research article are given below:
 \begin{itemize}
     \item A two stage model is proposed for LV segmentation
     \item A new scheme of parameter optimization is presented.
     \item A model is proposed which is robust to the intensity inhomogeneity and shape variability of the endocardium.
 \end{itemize}

\section{LITERATURE REVIEW} \label{s:back}

Accurate segmentation of the LV holds paramount importance in the field of cardiology. It serves as the cornerstone for advanced diagnostics and treatment planning. Many researchers have invested critical time and effort in improving the accuracy and efficiency of LV segmentation \cite{tayebi2023automated,shoaib2023overview,tamoor2021automatic}. Table \ref{tab:myfirstlongtable} shows some of recent pivotal studies and contributions in the field of LV segmentation.


 


\begin{table}
\centering
\caption{Related Studies} 

\label{tab:myfirstlongtable}

\begin{tblr}{
  row{1} = {fg=red},
  hlines,
  vlines,
}
  Research
        & 
  Method
                                                                            & 
  Dataset
                         \\
Wang et al, 2023    & Lightweight ECA-Residual
  module                                                     & Synapse                             \\
Deepa et al, 2023   & 3D diffeomorphic algorithm                                                            & ACDC                                \\
Hong et al, 2022    & Dual encoder network                                                                  & ACDC                                \\
Awasthi et al, 2022 & LVNet                                                                                 & Cardiac US image of the
  canine LV \\
Dina et al, 2022    & UNet                                                                                  & ACDC and LVSC                       \\
Maria et al,2021    & ACM                                                                                   & York, Sunnybrook and ACDC           \\
Bi et al, 2021      & ACM                                                                                   & MICCAI and Shuo Dataset             \\
Irshad et al,2021   & Morphological tuning and
  active contours                                            & Sunnybrook                          \\
Kushbu et al, 2021  & {
  Watershed algorithm and Global Local
  \\Region Based Active Contour
  method
  } & 24 CMRI images                      \\
Zou et al, 2021     & Deep learning and
  curriculum learning                                               & Tongji Medical College of
  HUST    
\end{tblr}
\end{table}

 \cite{wang2023mafunet} has introduced a lightweight ECA-Residual module for the segmentation of the LV. This paper lays focus on the number of parameters that can be trained. It extracts the features while decreasing the number of parameters.  It also makes use of a spatial attention gating module to suppress irrelevant areas in an image. This guarantees the efficiency of the model while enhancing its performance simultaneously. The dataset used for this study was Synapse dataset.  It achieved a dice score of 0.84.

In another research Deepa et al \cite{krishnaswamy2023novel,naseer2022computer} aims to segmentation the LV using a 3-dimensional diffeomorphic algorithm. It calculates voxel-to-voxel correspondence in 3D space. This method permits the constraints to be implemented so that the deformation can be controlled for better accuracy. It uses a radial and three curl components as parameters. The dataset used includes the Automated Cardiac Diagnosis Challenge (ACDC) dataset and the Mazankowski Alberta Heart Institute’s Dataset. The entire datasets were not, however, utilized for this task. In summary, this method has shown advancing results in this field.

On the other hand Hong et al \cite{hong2023dual,naseer2024occupancy} proposed a dual encoder network that works along with a transformer-CNN architecture. The main idea of this research was to counter the limits of usual CNN and transformer based techniques. The combination of these two architectures has led to improvement in the accuracy in the final segmentations. This method uses two encoders. The first encoder is a Swin-transformer that captures global information. It makes use of a self-attention mechanism to model long-range dependencies. While the second ender is the CNN encoder. It captures the local information using convolutional operations. After this a fusion module is used to combine both the acquired features with the help of skip connections. The dataset used in this research for segmentation of the LV is the Automated Cardiac Diagnosis Challenge (ACDC) dataset. The results show a dice score of 0.91.

Awasthi et al \cite{awasthi2022lvnet} worked on a lightweight segmentation model known as the LV Network (LVNET). This research uses Ultrasound images for segmentation rather than CMR scans. This model utilizes fewer parameters while improving on the scores achieved by its predecessors. It sets the US images as an input while utilizing the segmented images as the output during training. The proposed model utilizes only 5\% of the memory in comparison to the UNet model. The dice score of this study was 0.86-0.88. This technique appears effective as it requires a lesser amount of memory to perform the segmentation. It was concluded if more tuning of hyperparamters will be done, the greater the likelihood to acquire improved results.

Research by Dina et al \cite{abdelrauof2022light-weight,raza2023alzheimer} aims at making a CNN based model. The localized module utilizes a lightweight model to calculate a region of interest (ROI). Additionally, a UNet model enhanced by a multi-Gate block is used. ACDC and Left Ventricle Segmentation Challenge are two datasets which are used in this research. The dice score achieved by this was 0.9 and 0.91 on the respective datasets. As UNet model is a deep learning (DL) technique, the cost of DL is overly expensive  for training. 

Active Contour Model (ACM) is used by Tamoor et al \cite{tamoor2021two-stage}, proposing a two-stage model for robust segmentation of the LV. This research utilizes statistical methods for capturing image intensity, region boundary calculations with edge-based terms, regularization functions, etc. The first of the two stages include thresholding, removing boundary artifacts. The second stage transmits the image through an ACM. 
This study has presented promising results with a high dice score on one of the latest datasets. However, same set of parameters are used across all different types of MRI slices.

While Ke et al \cite{bi2021sequential} also uses ACM, alongside with Gradient Vector Convolution (GVC). This research has concentrated on a sequence of CMR images. The images in a 3-dimentional CMR are checked with the images in the neighbor, ensuring that the segmented images are similar. Sequential Shape Similarity (SSS) was utilized to determine how similar these shapes in the images are. This research successfully showed a Mean Absolute Distance (MAD) of 1.15 m. 

Mehreen et al \cite{irshad2021discrete,malik2022applying,chughtai2023content}, presents a method that morphological tuning and active contours (MTAC) for segmenting LV. This approach has four stages which make it computationally expensive. 
The first stage utilizes automated localization. The second stage involves preparing the images to enhance their quality. The third stage utilizes the MTAC algorithm to segment the images. Lastly, contrast-based refining helps to solve some minor issues with boundaries. 

While Kushbu et al \cite{kushbu2021interactive}, uses a hybrid method for the segmentation of the LV. Watershed and Global Local Region Based Active Contour are created to form the proposed model. 
The results show a mean HD of 0.927. The reduction in processing time is done by selecting a ROI. This can be an effective idea. However, it must be considered that the time for selection of ROI is worth it. If the ROI selection is too long, then this option may not be worth it.

The accurate segmentation of the LV alongside with DL and curriculum learning was incorporated by Zou et al \cite{zou2021novel}. Initially SinMod is used to locate the ROI. Then the U-Net is used to get the segmentation of the LV. 
This process utilizes DL that is computationally expensive. Moreover, this additional layer of curriculum learning is applied, so this may be a lengthy process.

Apart from these Degerli et al \cite{degerli2021early} introduced a three-phase technique for segmentation of LV. The initial phase focuses on the segmenting of LV using a DL. While the second phase is focused on analyzing the segmentation with the assistance of feature engineering. The final phase will determine the presence of Myocardial infarction, MI (heart attack). The HMC-QU dataset is used for this research. It manages to give a sensitivity of 95\%. The results may be satisfactory, but the issue will arise at the expense of computational resources.

Based on previous study, a few research gaps can be identified. For accurate LV segmentation, structural Variations in the different slices of one MRI is not easy to handle by one set of parameters,  such as differences in shape and size between basal, mid-ventricle, and apical regions. Research could focus on developing techniques that effectively handle these variations to ensure accurate segmentation across the entire LV volume. Second, another challenge is to handle the dynamic nature of the heart across different phases of the cardiac cycle (diastolic and systolic). Existing methods may not adequately account for these variations, leading to suboptimal segmentation results. 

\section{RESEARCH METHODLOGY} \label{s:sec2}
A two-phase model is proposed for segmentation of the LV. ACDC dataset is utilized in this research, which contains 3-D CMR scans of the patients. These 3-D scans are then transformed to multiple 2-D slices. The CMR scan of a single patient can have from 6-16 (2-D slices). The two-phase model passes these 2-D slices to acquire the final segmentation. The slices will be classified, and a binary mask will be created and based on the results of these two the parameters for the final segmentation will be determined. The two phases of this model are as follows:

Phase 1: Selection of an optimized set of parameters based on the type of slice.

Phase 2: Segment LV with the parameters defined in phase 1.

\begin{figure}[h]
    \centering   \includegraphics[width=0.7\linewidth]{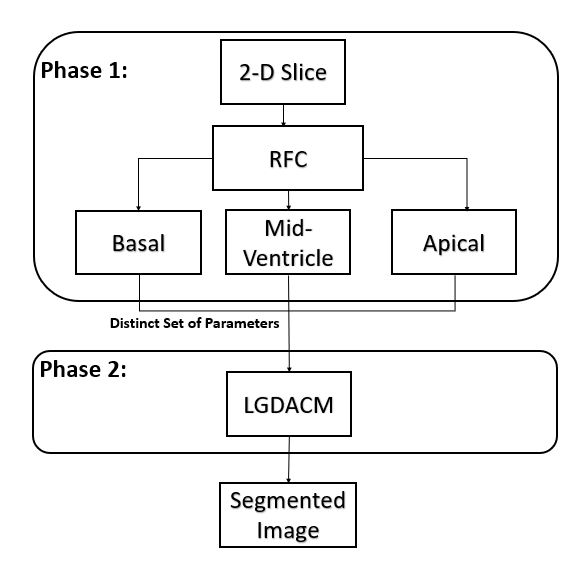}
    \caption{2-Stage Mode}
    \label{fig:2-Stage Mode}
\end{figure}

As shown in figure \ref{fig:2-Stage Mode}, the proposed method is a two-stage model for accurate segmentation of LV. The initial phase involves a 2-D slice of the CMR being classified as either Basal, Mid-Ventricle, or Apical. The Random Forest Classifier (RFC) algorithm is employed to classify these images into three classes. Some of the images are manually labeled by experts to create a training set and validation set. As the term suggests, training data will be used to train the machine learning model (RF algorithm) for classification. On the other hand, the validation dataset ensures that the model is performing accurately on the unseen data as well. Based on how the slice is classified separate set parameters will be assigned for the second phase.

\subsection{Dataset}
The dataset used for this research is the Automated Cardiac Diagnosis Challenge (ACDC) Dataset \cite{bernard2018deep}. The ACDC dataset is collected through a medical examination conducted at the University Hospital of Dijon, located in France. It took six years to gather all the data for this dataset, utilizing two MRI scanners during this period. The ACDC dataset comprises Cardiac MRI (CMR) scans from a total of one hundred patients in the training set. Each CMR is accompanied by its corresponding ground truth data. The ACDC dataset is recognized as one of the most recent and up to date CMR Benchmark Datasets for LV segmentation tasks. The dataset comprises 150 examinations, each originating from distinct patients and categorized into five evenly distributed subgroups. These subgroups consist of four pathological groups and one group representing healthy subjects (30 NOR subjects, 30 patients with previous myocardial infarction (MI), 30 patients with hypertrophic cardiomyopathy (HC), and 30 patients with abnormal right ventricle (ARV). The data were acquired over a span of six years utilizing two MRI scanners with varying magnetic strengths (1.5 T - Siemens Area, Siemens Medical Solutions, Germany, and 3.0 T - Siemens Trio Tim, Siemens Medical Solutions, Germany). Cine MR images were captured using a breath-hold technique, employing either retrospective or prospective gating, and employing a SSFP sequence in the short-axis orientation. 28 to 40 volumes are obtained per subject, to introduce natural variability in image quality and cardiac conditions. 

The ACDC dataset serves as a valuable resource for training and evaluating LV segmentation models. Its recentness and comprehensive nature make it a suitable choice for conducting research in the field of cardiac image analysis.

\subsection{First phase of the proposed method}
\begin{figure}
    \centering   \includegraphics[width=0.7\linewidth]{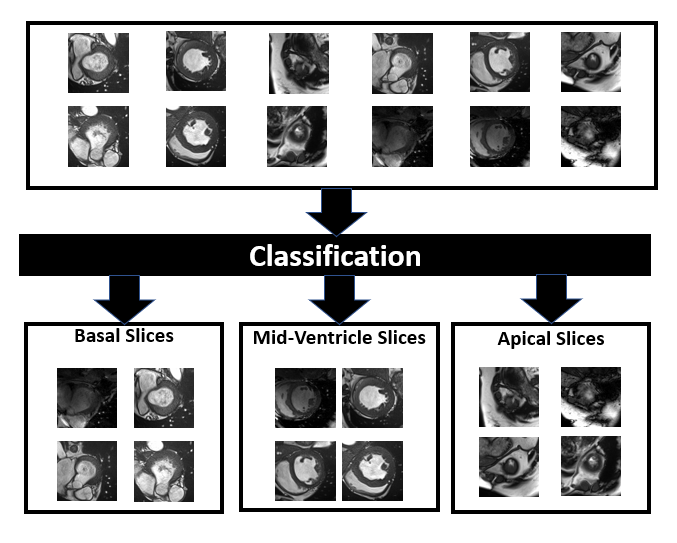}
    \caption{Classification}
    \label{fig:Classification}
\end{figure}

This first phase, as shown in figure \ref{fig:Classification}, is classification of the images into one of the three distinct classes. To create a training and testing dataset, some of the datasets are labeled manually. The images features are obtained from Daisy Features. To classify images into the three classes the ML algorithm known as Random Forest is used. The following steps are used to classify the CMR images.
\begin{itemize}
  \item Manually define labels.
  \item Extract DAISY Features.
  \item Inverse Position Index
  \item Shuffling the labeled data
  \item Classification using Random Forest
\end{itemize}
\subsubsection{Manually Define Labels}
20 CMRs of different subjects are chosen for labelling. Every fifth patient is selected from the dataset to be labeled to have unbiased selection. As the patients in the dataset are divided into 5 categories as explained above. The first twenty patients are of the first category and similarly all the categories have twenty patients each respectively. Having picked every fifth patient ensures that there will remain balance in the division of data from the dataset among all the categories.
\subsubsection{Extract DAISY Features}
Some image features are required to classify them. For this a robust feature extraction technique is required. The technique used in this research is Descriptor for Various Aggregated Image Descriptor (DAISY Features). Engin Tola et al \cite{yao2009parallel} introduced the world to the DAISY Features. They are local feature descriptors with a center-symmetrical structure. The daisy features utilize different directional diagrams of an image using some Gaussian filters of different sizes. It contains Rings that are overlayed on the different patches of the image, this arrangement increases its robustness \cite{ghorbani2022novel}. It calculates the descriptors from the patches of a local image and outputs features that represent the entire image. It is commonly used for tasks like image recognition, analysis of textures and detection of objects in an image. This method is known to be very efficient \cite{li2014image}.

To extract the daisy features, the raw 2D slices underwent processing through various techniques such as intensity adjustment, histogram equalization, median blur, etc. However, the most precise outcomes were seen when the DAISY feature extractor was applied directly to the 2D slices.  However, it's noteworthy that all images were resized to dimensions of 200 by 200.
\subsubsection{Inverse Position Index}
Following the calculation of the DAISY features, the next step is to add an additional value to them. The Inverse Position Index (IPI) is the new value. This term has a value that represents the position of a slice in the 3D CMR. The term can be calculated as:
\begin{equation}
\frac{n-p+1}{n}
\end{equation}
Where p denotes the position of the slice and n is the total amount of slices. The initial slice is always allocated the IPI value 1 regardless of the number of slices. The IPI value always decreases as move to the next slice. The decrease will be higher in the case of small number of slices while the rate of decrease will be smaller for the ones with higher number of slices.

\subsubsection{Train-Test-Split}
After the features have been acquired the next step is to shuffle the data. To shuffle the data, a built-in python function known as train-test-split is utilized. As it shuffles data and smartly divides into two splits. Train-test-split divides the data into two groups. These two parts are train split and test split. It also takes steps to ensure that the ratio of the labels is similar in both parts. It will also ensure that similar features will be divided with equal ratios in both the parts. Determining its ratio can be of great significance. As the size of this split can have enormous effect of the training and outcome  \cite{racz2021effect}. This technique is widely used by almost all ML algorithms \cite{tan2021critical}.

In this step both the splits will have similar ratios of similar data and labels. During this stage, both partitions will exhibit comparable ratios of analogous data and labels. The training-test split allocates 80\% to training and 20\% to testing (80 subjects (16 per group) for training and 20 for validation), with the incorporation of a random seed to guarantee consistent outcomes across multiple runs. 50 subjects are used for testing. The use of a random seed ensures that the data is consistently partitioned in the same manner whenever the train-test-split is executed.

\subsubsection{Random Forest for classification}
An extremely important part of the classification is the use of an ML algorithm to get final labels. The ML model used for this study is the Random Forest Classifier (RFC). RFC is an ML algorithm that is developed with the assistance of numerous decision trees. While a single decision tree presents some rules for the classification process \cite{seeja2023novel}. The decision tree connects different nodes to each other via branches. One node is a statement that takes us to the next node (one node will be selected from multiple nodes) until a leaf node is reached. The leaf node will be assigned a class that can be defined as the result of the tree. There are multiple leaf nodes and each one will have a class attached to it \cite{joshuva2019comparative}. The term forest in Random Forest Classifier algorithm, as the name suggests, is like a common forest, which means it will tend to have many trees in it. While the term random suggests that it will pick some random entries and random features to create a single decision tree. This process will be repeated to create numerous decision trees. All decision trees will have their own results. All these results will be passed forward and the most frequently encountered answer will be taken as the actual result.

As RFC model is based on randomly selecting entities and attributes to form a new subset. So, we once more will have the need to use random seed. If will ensure that the RFC will run in the exact same way every single time. 

\begin{figure}[h]
    \centering
    \includegraphics[width=0.8\linewidth]{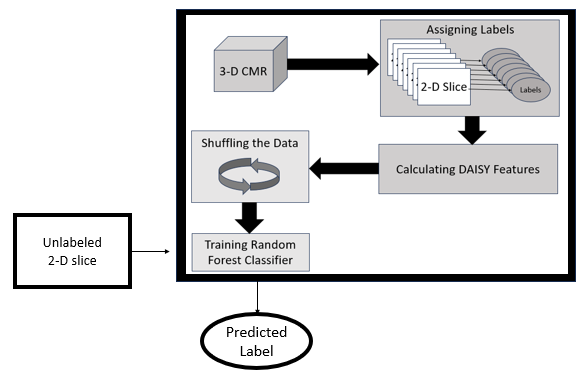}
    \caption{Phase-1}
    \label{fig:P1}
\end{figure}

After the training phase is completed, the model will be ready to select the 2-D slices into one of the three categories. Figure \ref{fig:P1} shows how the 2D slice will have a label assigned to it.

\subsection{Second phase of Proposed method}
In the next phase, the creation of a mask is performed. The LGDACM will create a 2-D segmented image using a mask and a set of parameters.
\subsubsection{Creating a Mask}
The mask plays a crucial role in achieving a final segmentation. The better the mask will be, the better the segmentation. The process of creating a mask is aided by regionprops. It extracts the properties of a region in an image \cite{mondal2015image}. There are many regional properties used in this research. Such as Area, eccentricity, centroids, etc.

Initially, to compute the mask, we employ an image intensity adjustment technique across all images. The intensity of an image refers to the mean of all the gray-scale pixels in an image. This adjustment aids in enhancing performance across various criteria, such as edge detection \cite{montazeri2016light}. The adjustment is based on the current intensity of each image, determined by the formula provided below. The formula for intensity adjustment is:

\begin{equation}
    \alpha = \frac{100}{2 \cdot x}
\end{equation}
\begin{equation}
{Image Intensity} = (\text{image} \cdot \alpha) + \beta
\end{equation}

\begin{figure}
    \centering
    \includegraphics[width=\linewidth]{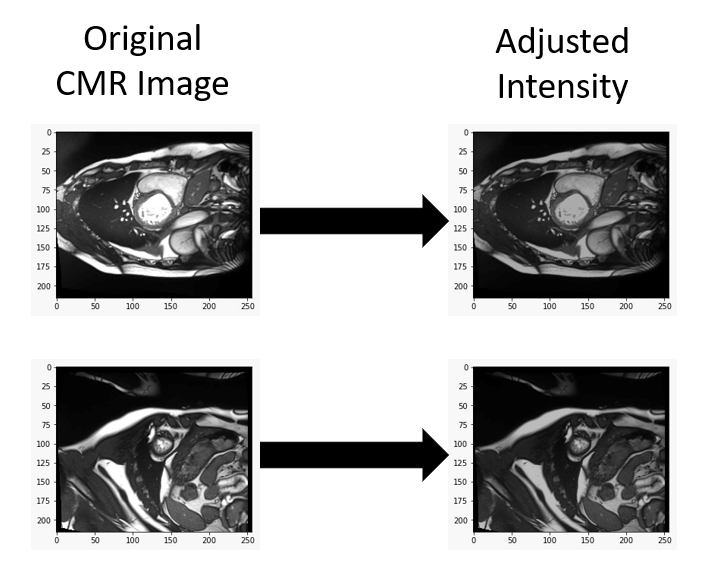}
    \caption{Intensity Adjustment}
    \label{fig:adjusted}
\end{figure}

In the given equations, $\alpha$ represents alpha, $x$ denotes the intensity of the current image, and $\beta$ stands for beta. As depicted in the given equation, the value of $\alpha$ is contingent upon the intensity of the image being analyzed, while the initial value of $\beta$ is zero. If any issues arise during mask creation, the value of beta will be adjusted in subsequent steps, leading to a recalibration of image intensity. Figure \ref{fig:adjusted} a comparison is presented between the original CMR image and its counterpart after intensity adjustment. In images without intensity adjustment, the left ventricle (LV) is undetected in 6\% of cases, resulting in a dice score of 0.72 when comparing the created masks to the corresponding ground truths. Conversely, when intensity adjustment is applied, the LV is not found in only 2.7\% of images, and the dice score increases to 0.77.

The method of mask creation in this study is based on the use of regionprops and a sequential process. The sequential process since the 2-D slices in a sequence in the 3-D CMR are going to be like the previous slice in terms of location \cite{bi2021sequential}. Initially, only the initial three slices are considered to calculate the binary masks. For a single mask, optimal threshold value is determined among multiple threshold values based on properties, like area, centroid, distance, circle. Prior information about the stucture of LV is incorporated that the LV mask is located nearer to the center and normally follows a circular shapre except basal slice \cite{dharanibai2014automatic}. Now the top three masks are there, and it is ensured that they are around the same location. This means that the centroids of all three LV masks must have minimal distance from one another. Since all the slices of the CMR will be around the same location \cite{bi2021sequential}.

\subsubsection{LGDACM}
Local Gaussian Distribution Active Contour Model (LGDACM) is an algorithm used in the segmentation of images. LGDACM is introduced by Li Wang et al \cite{wang2009active}. It makes use of means, variances, and Gaussian distributions to segment an image. The technique uses a level set function to fit a Gaussian distribution. The mean and variance will be derived as variables \cite{wang2009active}. It is well known for management of noises in an image. It can differentiate between regions with alike intensities but different variances. 

\subsubsection{Segmentation of LV Slices}

The LV is segmented using the LGDACM segmentation algorithm. There are three separate sets of parameters for the three distinct slices. 

\subsubsection{Basal Slices}
The parameters of the LGDACM are run in loop. This ensured that the finest set of parameters will be designated for the segmentation of basal slice of the LV. Similarly masks of different sizes are used as well to get the best segmentation results in the end. The 2D slice, the initial contour and a set of parameters will then be sent to the LGDACM for segmentation. The timestep \(\tau\) is set as 0.05 as a smaller step will ensure that the edges are not missed due to a large step. The large timestep can be time saving but may cost us in the accuracy \cite{li2005level}. While lambda1 $(\lambda)_1$ and lambda2 $(\lambda)_2$ are set to 3 and 2 respectively. Nu \(\nu\) was set to 0.0008*255*255.

\subsubsection{Mid-Ventricle Slice}
\begin{figure}
    \centering
    \includegraphics[width=\linewidth]{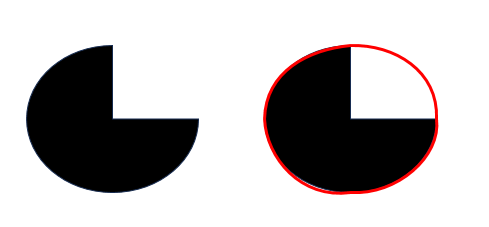}
    \caption{Convex Hull}
    \label{fig:Hull}
\end{figure}

Like the basal slices the parameters of the LGDACM are run in loop to ensure that the most well performing set of parameters will be picked for the segmenting the LV. Similarly masks of different sizes are used again to get the right segmentation results in the end. The 2D slice, the initial contour and the second set of parameters will then be sent to the LGDACM for segmentation. A change can occur depending on a 10-pixel radius from the midpoint of the mask. Standard deviation of the values will be calculated. If the radius has only the LV inside, then the pixel values will be similar, and the standard deviation will be small. The size of LV in this case can also be seen as large. The smaller the LV the less will be the standard deviation. The number of iterations will depend on the standard deviation. The contour will evolve based on the size of the LV. The timestep \(\tau\) is set as 0.05 as a smaller step will ensure that the edges are not missed due to a large step. While lambda1 \(\lambda\)1 and lambda2 \(\lambda\)2 are set to 3.5 and 2.5 respectively. Nu \(\nu\) was set to 0.0005* 255*255 as it as reported by Wang et al \cite{wang2009active} it is the best value for Nu when dealing with the cardiac images. The experiments also confirm this as the results had a 2\% improvement. After the segmentation process was completed, there is a papillary muscle in the mid-ventricle. This muscle must be a part of the segmentation. But the LGDACM fails to add it to the segmentation. To counter this, use of the convex hull technique is incorporated. The convex region is a region that has a no open spaces. If a region has any open spaces, then it will be shut off by the convex hull. Figure \ref{fig:Hull} demonstrates the convex hull. In this figure the left region is a concave shape while the region on the right is the convex hull of the concave region as it draws a line that wraps it filling up the empty region. Using this algorithm, it is managed to add the papillary muscle to the segmentation.

\subsubsection{Segmentation of Apical}
These slices are often very small in size and blood pool is negligible that they are not visible to the naked human eye. Just like the basal and mid ventricle slices, the apical slice is also segmented by running a loop with different set of parameters and masks to ensure best possible results of segmentation.

The shrinkage of the mask will go through two loops like the mid-ventricle. But there is one difference that if the masks have less than 120 pixels in it then in that case the process of shrinking will be stopped. Now this shrunk mask along with the 2D slice will be passed to the LGDACM in octave. The number of iterations for this slice naturally will be very small. As the size of apical is small so the needed evolution will also be small. The timestep \(\tau\), Mu \(\mu\) and Nu \(\nu\) will be same as the one for mid-ventricle. While lambda1 \(\lambda\)1 and lambda2 \(\lambda\)2 are set to 1.75 and 1.5 respectively.

\begin{figure} [h!]
    \centering
    \includegraphics[width=0.8\linewidth]{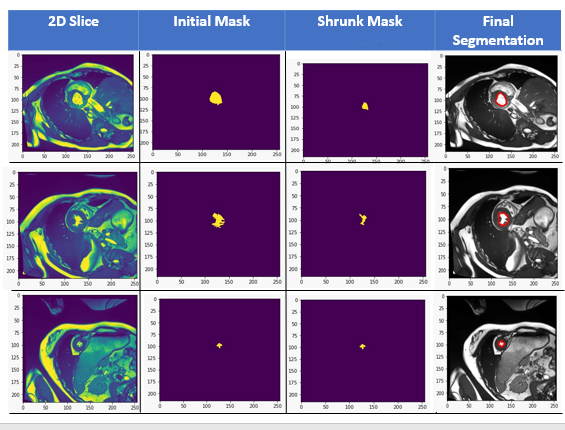}
    \caption{Phase-2}
    \label{fig:ip}
\end{figure}

{Figure} \ref{fig:ip} shows the process the 2D slices going through for the segmentation of the image. The first column has the 2D slices. While the second column represents their binary masks. The third one is the shrunk form of these masks. And the last column has the outcomes of our segmentation alongside with boundaries of the LV being marked out. The rows show an example image from the basal, mid-ventricle and apical classes respectively

\begin{figure} [h!]
    \centering
    \includegraphics[width=0.7\linewidth]{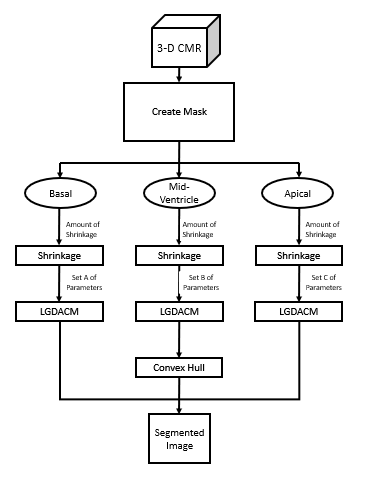}
    \caption{Phase-2}
    \label{fig:P2}
\end{figure}

Figure \ref{fig:P2} shows the stepwise approach of phase 2. At the start a mask is created and based on the label it will go on one of the three paths. After this shrinkage is applied. After that LGDACM is used to segment the basal, mid-ventricle and apical, with a separate set of parameters. And finally, the mid-ventricle will have the convex hull applied to it.

\section{Results}

\subsection{Results for First Phase}
To evaluate the performance of the first phase, 10-fold cross validation is used. This method is considered to be a highly recommended evaluation matric for classification models \cite{kohavi1995study,zhang2015cross}. It is also known to have a very small bias in comparison with other evaluation techniques. The following table \ref{tab:2} shows the accuracies for the all the fold along with the mean score:

\begin{table}[H]
    \centering
    \begin{tabular}{ccccccccccc}
 \hline

  \hline
  Folds& 1& 2& 3& 4& 5& 6& 7& 8& 9& 10  \\ 
  \hline
      Accuracy	&0.902	&0.976	&0.9	&0.925	&0.925	&0.925	&0.95	&0.925	&0.85	&0.95  \\
  \hline
    \end{tabular}
    \caption{Results for K-Fold Cross Validation}
    \label{tab:2}
\end{table}

Table \ref{tab:2} shows the accuracies achieved by all the folds. The mean (Average) of all the accuracies is 0.9228 or 92.28\%. A test-train split of 0.2 was created to divide the data into two sets. Using the RFC model the following results are generated on the test data:

\begin{table}[H]
    \centering
    \begin{tabular}{cccc}
 \hline

  \hline
  	&Precision (P)	&Recall (R)	&F1 Measure  \\ 
  \hline
    Basal&	0.94&	0.91&	0.92\\
    Mid-Ventricle&	0.89&	0.94&	0.92\\
    Apical&	1.00&	0.92&	0.96\\
  \hline
    \end{tabular}
    \caption{Evaluation of the Classification}
    \label{tab:3}
\end{table}

Table \ref{tab:3} shows the classification report of the classified data. These measures are well known to handle the class imbalance and give accurate results \cite{hand2017note}. These measures can be calculated as:
\begin{equation}
    Precision = P(\text{True Match}\,|\, \text{Predicted})
\end{equation}
\begin{equation}
    Recall = P(\text{True Match}\,|\, \text{Actual})
\end{equation}
\begin{equation}
    F1 = \frac{2 \cdot \text{Precision} \cdot \text{Recall}}{\text{Precision} + \text{Recall}}
\end{equation}

Recall can be defined as the correctly assigned labels to a class by all the labels predicted to the class. While precision is correctly assigned labels to a class out of the actual labels. F1 uses both precision and recall calculating an optimal score in place of an average. If the data is not balanced, then average is not going to be a good measure, hence the F1 score is used \cite{chicco2020advantages}. Almost all scores are over 90\% which show that the results of the classification phase is very promising.

\subsection{Results for Second Phase}
The evaluation of the second phase's performance relies on the Dice Score Coefficient (DSC). Table \ref{tab:4} displays the Dice scores for the segmentation process. To facilitate a thorough comparison and highlight the efficacy of the proposed method, results are presented with varied parameter settings. In the first row of Table \ref{tab:4}, the proposed setting involves employing a distinct optimized parameter set for each type of cardiac MR slice. Conversely, the subsequent three rows showcase results obtained using a uniform parameter setting for all types of slices.

\begin{figure}
    \centering
    \includegraphics[width=0.5\linewidth]{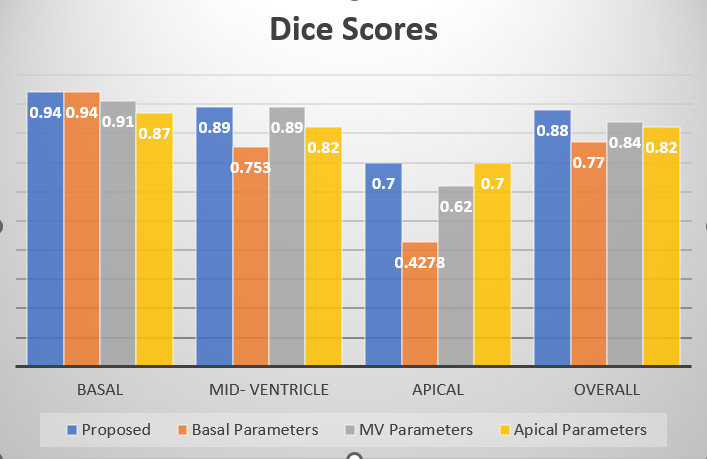}
    \caption{Dice Score}
    \label{fig:Dscore}
\end{figure}

In table \ref{tab:4} and figure \ref{fig:Dscore} , the overall Dice score achieved by the proposed method is 0.88. Through the utilization of parameter optimization scheme, the outcomes produced demonstrate a superiority compared to those generated by employing a uniform set of parameters across all slices. Table \ref{tab:4} delineates parameter sets A, B, and C, corresponding to the parameters applied for basal, mid-ventricle, and apical regions, respectively. Notably, employing a singular set of parameters reveals less favorable scores in contrast to the suggested approach. The optimal results achieved for one slice are found to be insufficient when applied uniformly across the remaining slices.

\begin{table}[H]
    \centering
    \begin{tabular}{ccccc}
 \hline

  \hline
  Set of Parameters&	Basal&	Mid- Ventricle&	Apical&	Overall  \\ 
  \hline
  Proposed&	0.94&	0.89&	0.70&	0.88\\
Parameter Set A&	0.94&	0.75&	0.43&	0.77\\
Parameter Set B&	0.91&	0.89&	0.62&	0.84\\
Parameter Set C&	0.87&	0.82&	0.70&	0.82\\

  \hline
    \end{tabular}
    \caption{Dice Scores}
    \label{tab:4}
\end{table}

Besides DSC, there are other evaluation techniques to evaluate the performance of the segmentation model. Some popular metrics include Jaccard Index (JI), F-measure, Hausdorff Distance (HD), Mean Average Distance (MAD) and Boundary Displacement Error \cite{wang2020image}. Table \ref{tab:5} given below presents some of the results of these measures.

\begin{table}[H]
    \centering
    \begin{tabular}{ccccc}
 \hline

  \hline
  Measure&	Basal&	Mid- Ventricle&	Apical&	Overall\\ 
  \hline 
Jaccard Index (JI)&	0.90&	0.84&	0.65&	0.8\\
Precision&	0.98&	0.92&	0.71&	0.91\\
Recall / Sensitivity&	0.91&	0.89&	0.80&	0.88\\
F-1&	0.94&	0.89&	0.70&	0.88\\
Accuracy&	0.99&	0.99&	0.97&	0.99\\
Hausdorff Distance (HD)&	2.12&	2.23&	1.93&	2.12\\
Mean Average Distance (MAD)&	1.54&	2.07&	3.68&	2.1\\
Mean Absolute Error (MAE)&	0.26&	0.26&	0.2&	2.3\\
Specificity&	0.999&	0.99&	0.97&	0.99\\
Boundary Displacement Error (BDE)&	0.13& 0.15& 1.59& 0.35\\
  \hline
    \end{tabular}
    \caption{Evaluation Scores}
    \label{tab:5}
\end{table}

\section{CONCLUSIONS \& FUTURE WORK}
The attention of this proposed method is on the segmentation of the LV based on the types of LV slices. The whole LV is classified into 3 parts. These parts are basal, mid-ventricle and apical. Using CMR scans, the input is a single 3D scan of the LV which is further divided into multiple 2D slices. A new approach is propose to define different optimized set of parameters for each type of CMR slice. To perform the segmentation, a two-phased approach is used. Where the first phase is concerned with classifying the 2D slices into basal, mid-ventricle and apical class. While the second phase focuses on segmenting the 2D slices based on the results obtained from the initial phase.

\begin{figure} [h!]
    \centering
    \includegraphics[width=0.8\linewidth]{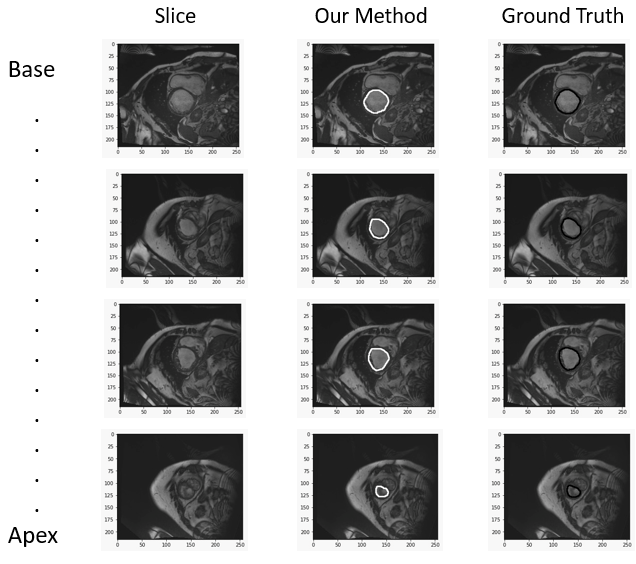}
    \caption{Our Method and Ground Truth}
    \label{fig:sgs}
\end{figure}

The comparison between our method's acquired results and the ground truth is depicted alongside the 2D slices in Figure \ref{fig:sgs}.

By harnessing the ability of CMR scans and ML, a two-phase approach for the accurate segmentation of LV is presented. The research shows that the separate set of parameters based on the type of LV slice can be very effective. The set of parameters set for the basal are optimal for the basal slice but suboptimal for apical and mid-ventricle and vice versa. This problem holds for all existing state-of-the-art techniques used for the segmentation techniques, Active contour model (snake model) and its variation suffer from the same issue, in medical image segmentation. Onet set of width of kernel, internal, external forces is not optimal to accurately evolve the curve at the boundary of ROI for different types of slices of CMR \cite{wang2009active,tamoor2021two-stage}. In deep learning models and other machien learning models they also suffer with parameter optimization problem to handle variation in the size and shape of LV slices.\cite{tamoor2021two-stage,hajiaghayi20183d,queiros2014fast} The proposed model that uses separate sets of parameters outperforms the single set parameter approach. The results of the experiments demonstrate that the initial phase managed to achieve a mean score (with K-Fold Cross Validation) of 0.9228 using RFC. In the second phase, an overall score archived by DSC was 0.88, with LGDACM. While the DSC score for basal, mid-ventricle, and apical was 0.94, 0.89, and 0.70 respectively.

A few limitations of the proposed model can be listed which will give researchers new directions in LV segmentation to improve results. First, the model relies on manually labeling a subset of the dataset for training and validation purposes. Manual labeling can be time-consuming, subjective, and prone to human error, potentially limiting the scalability and generalizability of the model. Second, the model's performance may be limited to the specific characteristics of the ACDC dataset used in the research. More usuability of this method needs to be explored in future. This Proposed model consists of two stages so implementing and deploying such a model in real-world clinical environments may require robust infrastructure and computational resources.

In future, more parameter optimization methods can be used to further improve the results and alternative techniques for creating masks and multiple algorithms also can be incorporated into the second phase. Proposed method can be used with CNN models and other classifications methods. Given the continuous pursuit of better and innovative solutions in the field of medicine, segmentation algorithms have considerable promise. In the foreseeable future this methodology may usher in a new era of advancements, contributing to cardiac healthcare.

\bibliographystyle{unsrt}  






\bibliography{Cardiac.bib}

\end{document}